\documentclass[12pt]{article}

\usepackage{ a4, graphicx, amsmath, amsthm, color, ulem, psfrag, url } 

\newcommand{ \dn }[ 1 ]{ \boldsymbol #1 }
\newcommand{ \mt }[ 1 ]{ \mathrm #1 }

\newcommand{ \be }{ \begin{equation} }
\newcommand{ \ee }{ \end{equation} }
\newcommand{ \bmath }{ \begin{displaymath} }
\newcommand{ \emath }{ \end{displaymath} }

\newcommand{ \by }{ \dn{ y } }

\newcommand{ \ident }{ \mt{ I }_{ n } }
\newcommand{ \identst }{ \mt{ I }_{ \nstar } }
\newcommand{ \nstar }{ n^* }

\newcommand{ \bbl }{ \dn{ \beta }_{ \ell } }
\newcommand{ \bbo }{ \dn{ \beta }_{ 0 } }

\newcommand{ \bx }{ \mt{ X } }

\newcommand{ \bxl }{ \bx_{\ell} }
\newcommand{ \bxo }{ {\bx}_{0} }

\newcommand{ \doS } { {\mt{\Sigma}_\ell'} } 

\newcommand{ \bboup }{ \overline{\overline{ \dn{ \beta }} }_{ 0 } }

\newcommand{ \varl }{ \sigma_\ell^2  }
\newcommand{ \varo }{ \sigma_0^2  }

\newcommand{ \xtxl }{ \big( \mt{X}_\ell^T \mt{X}_\ell \big) }

\newcommand{ \invxtxl }{ \xtxl^{-1} }

\newcommand{ \xtxstarl }{ \big( {\mt{X}_\ell^{*}}^T \mt{X}_\ell^* \big) }

\newcommand{ \invxtxstarl }{ \xtxstarl^{-1} }

\newtheoremstyle{mytheoremstyle}{3pt}{3pt}{\itshape}{}{\scshape}{:}{0.5em}{}
\theoremstyle{mytheoremstyle}

\newtheorem{lemma}{Lemma}

\usepackage[comma,longnamesfirst]{natbib}

\begin{document}

\normalem

\title{\textbf{Limiting Behavior of the Jeffreys Power-Expected-Posterior Bayes Factor in Gaussian Linear Models}}

\author{
D.~Fouskakis\thanks{D.~Fouskakis is with the Department of Mathematics,
National Technical University of Athens, Zografou Campus, Athens 15780
Greece; email \texttt{fouskakis@math.ntua.gr}}, \ and
I.~Ntzoufras\thanks{I.~Ntzoufras is with the Department of Statistics,
Athens University of Economics and Business, 76 Patision Street, Athens
10434 Greece; email \texttt{ntzoufras@aueb.gr}} 
}


\date{}

\maketitle

\maketitle

\noindent
\textbf{Summary:}
Expected-posterior priors (EPPs) have been proved to be extremely useful for testing hypotheses on the regression coefficients of normal linear models.
One of the advantages of using EPPs is that impropriety of baseline priors
causes no indeterminacy in the computation of Bayes factors. However, in regression problems, 
they are based on one or more \textit{training samples}, that could influence the resulting posterior distribution. on the other hand, the \textit{power-expected-posterior priors} are minimally-informative priors that reduce the effect of training samples on the EPP approach, by combining
ideas from the power-prior and unit-information-prior methodologies. In this paper we prove the consistency of the Bayes
factors when using the power-expected-posterior priors, with the independence Jeffreys as a baseline prior, for normal linear models, under very mild conditions on the design matrix.

\vspace*{0.15in}


\section{Introduction}

\cite{perez_berger_2002} developed priors for model comparison,
through utilization of the device of ``imaginary training samples"
\citep{good, spiegelhalter_smith_80, iwaki}. They defined the
expected-posterior prior (EPP) as the posterior distribution of a parameter
vector for the model under consideration, averaged over all possible
imaginary samples $\dn{ y }^*$ coming from a ``suitable" predictive
distribution $m^* ( \dn{ y }^* )$. Hence the EPP for the parameter vector $\dn{ \theta }_\ell~$, of any
model $M_\ell \in \cal{ M }$, with ${ \cal M }$ denoting the model space, is
\begin{equation} \label{epp} \pi^{EPP}_\ell ( \dn{ \theta }_\ell ) = \int
\pi_\ell^N ( \dn{ \theta }_\ell | \dn{ y }^* ) \, m^* ( \dn{ y }^* ) \, d
\dn{ y }^* \, , \end{equation} where $\pi_\ell^N ( \dn{ \theta }_\ell | \dn{
y }^* )$ is the posterior of $\dn{ \theta }_\ell$ for model $M_\ell$ using a
baseline prior $\pi_\ell^N ( \dn{ \theta }_\ell )$ and data $\dn{ y }^*$.

An attractive option for $m^*$ arises from selecting a ``reference" or ``base" model $M_0$ for
the training sample and defining $m^* ( \dn{ y }^* ) = m_0^N ( \dn{ y }^* )
\equiv f ( \dn{ y }^* | M_0)$ to be the prior predictive distribution,
evaluated at $\dn{ y }^*$, for the reference model $M_0$ under the baseline
prior $\pi_0^N ( \dn{ \theta }_0 )$.  For the variable-selection problem 
considered in this paper, the constant model (with no predictors) is used as a 
reference model, following the skeptical-prior approach
described by \citet[Section 5.5.2]{spiegelhalter_abrams_myles}. This selection simplifies computations, and makes the EPP approach equivalent 
to the arithmetic intrinsic Bayes factor approach of \cite{berger_pericchi_96a}.

One of the advantages of using EPPs is that impropriety of baseline priors
causes no indeterminacy in the computation of Bayes factors. With EPPs, we can use an 
improper baseline prior $\pi_\ell^N ( \dn{ \theta }_\ell )$ in (\ref{epp}), since the
arbitrary constants cancel out in the calculation of any Bayes factor. However, in regression problems, 
EPPs are based on one or more \textit{training samples}, that could influence the resulting posterior distribution.

To diminish the effect of training samples on the EPP approach and simultaneously to produce a minimally-informative prior, \cite{fouskakis_et.al_2014} 
introduced the \textit{power-expected-posterior} (PEP) priors, by combining
ideas from the power-prior approach of \cite{ibrahim_chen_2000} and the
unit-information-prior approach of \cite{kass_wasserman_95}. 
As a first step, the likelihoods involved in the EPP distribution are raised to the
power $1/\delta$ and then are density-normalized. This power
parameter $\delta$ is set equal to the size of the training sample $n^*$, to represent information equal to one
data point. Regarding the size of
the training sample, $n^*$, this is set equal to the sample size $n$; in this way 
the selection of a training sample and its effect on the
posterior model comparison is completely avoided. 

In what follows, we examine variable-selection problems in Gaussian
regression models. Thus, for any model $M_\ell$,
with parameters $\dn{ \theta }_\ell = ( \dn{ \beta }_\ell \, , \sigma_\ell^2 )$,
the likelihood is specified by
\begin{equation} \label{new2-1}
( \dn{ Y } | \mt{ X }_\ell, \dn{ \beta }_\ell, \sigma_\ell^2, M_\ell ) \sim
N_n ( \mt{ X }_\ell \, \dn{ \beta }_\ell \, , \sigma_\ell^2 \, \mt{ I }_n )
\, ,
\end{equation}
where $\dn{ Y } = ( Y_1, \dots, Y_n )$ is a vector containing the
(real-valued) responses for all subjects, $\mt{ X }_\ell$ is a $n \times
d_\ell$ design matrix containing the values of the explanatory variables in
its columns, $\mt{ I }_n$ is the $n \times n$ identity matrix, $\dn{ \beta
}_\ell$ is a vector of length $d_\ell$ summarizing the effects of the
covariates in model $M_\ell$ on the response $\dn{ Y }$ and $\sigma_\ell^2$
is the error variance. Furthermore, we denote the imaginary/training data set by $\dn{ y }^*$, their size by $n^*$,
and the corresponding imaginary design matrix by $\mt{ X }^*$ of size $n^* \times ( p + 1 ) \,$, where $p$
denotes the total number of available covariates. Following the PEP methodology we set $n^*=n$ and $\mt{ X }^* = \mt{ X }$, where $\mt{ X }$ is the original $n \times ( p + 1 ) \,$ design matrix.  

For any model $M_\ell \in { \cal M }$, we denote by $\pi_\ell^N ( \dn{ \beta
}_\ell, \sigma_\ell^2 | \bx_\ell^* )$ the baseline prior for model
parameters $\dn{ \beta }_\ell$ and $\sigma_\ell^2$, with $\bx_\ell^*$ being the imaginary design matrix under model $M_\ell$. Then the
\textit{power-expected-posterior} (PEP) prior, $\pi_\ell^{ PEP } ( \dn{ \beta
}_\ell, \sigma_\ell^2 | \, \bx_\ell^* \, , \delta )$, takes the following
form:
\begin{equation} \label{pep}
\pi_\ell^{ PEP } ( \dn{ \beta }_\ell, \sigma_\ell^2 \, | \, \bx_\ell^* \,
,\delta ) = \pi_\ell^N ( \dn{ \beta }_\ell, \sigma_\ell^2 | \bx_\ell^* )
\int \frac{ m_0^N ( \dn{ y }^* | \, \bx_0^* \, , \delta ) }{ m_\ell^N (
\dn{ y }^* | \, \bx_\ell^*\, , \delta ) } \, f ( \dn{ y }^* | \, \dn{ \beta
}_\ell \, , \sigma_\ell^2, M_\ell \, ; \mt{ X}_\ell^* \, , \delta ) \, d
\dn{ y }^* \, ,
\end{equation}
where $f ( \dn{ y }^* | \, \dn{ \beta }_\ell \, , \sigma_\ell^2, M_\ell \,
; \mt{ X }_\ell^* \, , \delta ) \propto f ( \dn{ y }^* | \dn{ \beta }_\ell
\, , \sigma_\ell^2, M_\ell \, ; \mt{ X }_\ell^* )^{ \frac{ 1 }{ \delta } }$
is the likelihood, evaluated at $\dn{ y }^*$, under model $M_\ell$, raised to the power of $1 / \delta$ and
density-normalized, i.e.,
\begin{eqnarray} \label{power_likelihood}
f ( \dn{ y }^* | \, \dn{ \beta }_\ell \, , \sigma_\ell^2, M_\ell \, ; \mt{
X }_\ell^* \, , \delta ) & = & \frac{ f ( \dn{ y }^* | \dn{ \beta }_\ell,
\sigma_\ell^2, M_\ell \, ; \mt{ X }_\ell^* )^{ \frac{ 1 }{ \delta } } }{
\int f ( \dn{ y }^* | \dn{ \beta }_\ell, \sigma_\ell^2, M_\ell \, ; \mt{ X
}_\ell^*)^{ \frac{ 1 }{ \delta } } d \dn{ y }^*} = \frac{ f_{ N_{ n^* } } (
\dn{ y }^* \, ; \, \mt{ X }_\ell^* \dn{ \beta }_\ell \, , \sigma_\ell^2 \,
\mt{ I }_{ n^* } )^{ \frac{ 1 }{ \delta } } }{ \int f_{ N_{ n^* } } ( \dn{
y }^* \, ; \, \mt{ X }_\ell^* \dn{ \beta }_\ell \, , \sigma_\ell^2 \, \mt{
I }_{ n^* } )^{ \frac{ 1 }{ \delta } } d \dn{ y }^* } \nonumber \\
& = & f_{ N_{ n^* } } ( \dn{ y }^* \, ; \, \mt{ X }_\ell^* \dn{ \beta
}_\ell \, , \delta \, \sigma_\ell^2 \mt{ I }_{ n^* } ) \, ;
\end{eqnarray}
here $f_{ N_d } ( \dn{ y } \, ; \, \dn{ \mu }, \dn{ \Sigma } )$ is the
density of the $d$-dimensional Normal distribution with mean $\dn{ \mu }$
and covariance matrix $\dn{ \Sigma }$, evaluated at $\dn{ y }$.

When the reference model $M_0$ is nested in all other models (like in our case) the EPP (and therefore the PEP prior) for the parameter vector under $M_0$ is clearly the same as the baseline prior, i.e.
$$
\pi_0^{ PEP } ( \dn{ \beta }_0, \sigma_0^2 \, | \, \bx_0^* \, ,\delta ) = \pi_0^{ N } ( \dn{ \beta }_0, \sigma_0^2| \, \bx_0^* ),
$$ 
with $\bx_0^*$ being the imaginary design matrix under model $M_0$.

The distribution $m_\ell^N ( \dn{ y }^* | \, \bx_\ell^* \, , \delta )$
appearing in (\ref{pep}) is the prior predictive distribution (or the
marginal likelihood), evaluated at $\dn{ y }^*$, of model $M_\ell~$, using the
power likelihood defined in (\ref{power_likelihood}), under the baseline
prior $\pi^N_\ell ( \dn{ \beta }_\ell, \sigma_\ell^2 \, | \, \bx_\ell^* )$,
i.e.,
\begin{equation} \label{new2-3}
m_\ell^N ( \dn{ y }^* | \, \bx_\ell^* \, , \delta )
= \int \! \! \int f_{ N_{ n^* } } ( \dn{ y }^* \, ; \, \mt{ X }_\ell^*
\dn{ \beta }_\ell \, , \delta \, \sigma_\ell^2 \, \mt{ I }_{ n^* } ) \,
\pi^N_\ell ( \dn{ \beta }_\ell \, , \sigma_\ell^2 \, | \, \bx_\ell^* ) \, d
\dn{ \beta }_\ell \, d \sigma_\ell^2 \, .
\end{equation}

Similarly, the distribution $m_0^N ( \dn{ y }^* | \, \bx_0^* \, , \delta )$
appearing in (\ref{pep}) is the prior predictive distribution, evaluated at $\dn{ y }^*$, of the reference model $M_0$, using the
power likelihood defined in (\ref{power_likelihood}) (with $\ell=0$), under the baseline
prior $\pi^N_0 ( \dn{ \beta }_0, \sigma_0^2 \, | \, \bx_0^* )$,
i.e.,
\begin{equation} \label{new2-3b}
m_0^N ( \dn{ y }^* | \, \bx_0^* \, , \delta )
= \int \! \! \int f_{ N_{ n^* } } ( \dn{ y }^* \, ; \, \mt{ X }_0^*
\dn{ \beta }_0 \, , \delta \, \sigma_0^2 \, \mt{ I }_{ n^* } ) \,
\pi^N_0 ( \dn{ \beta }_0 \, , \sigma_0^2 \, | \, \bx_0^* ) \, d
\dn{ \beta }_0 \, d \sigma_0^2 \, .
\end{equation}

Here we use the independence Jeffreys prior (or reference prior) as the
baseline prior distribution. Hence for any $M_\ell \in {\cal M}$ we have
\begin{equation} \label{JE}
\pi_\ell^N ( \dn{ \beta }_\ell \, , \sigma^2 \, | \, \bx_\ell^* ) = \frac{
c_\ell }{ \sigma_\ell^2 } \, ,
\end{equation}
where $c_\ell$ is an unknown normalizing constant; we refer to the resulting
PEP prior as J-PEP.

It is worth noting that our method, works in a
totally different fashion than fractional Bayes factors \citep{ohagan_95}. In the latter, a fraction $b$ of the full likelihood is used 
to ``properize" the baseline prior and the remaining fraction $(1-b)$ of the full likelihood is used for model comparison.  
In contrast, with our approach, the original
likelihood is used only once, for simultaneous variable selection and
posterior inference. Moreover, the fraction of the likelihood (power
likelihood) --- used in the expected-posterior expression of our prior
distribution --- refers solely to the imaginary data coming from a prior
predictive distribution based on the reference model.

\section{The conditional J-PEP prior distribution}

In the following, under any model $M_\ell$, we denote by 
$$
\mt{H}_\ell = \bxl \invxtxl \bxl^T \mbox{~and by~} \mt{P}_\ell = \ident - \mt{H}_\ell 
$$ 
and the corresponding measures based on $\bxl^*$ by $\mt{H}_\ell^*$ and $\mt{P}_\ell^*~$, respectively. 

Under (\ref{JE}), the corresponding marginal likelihood, with response data $\by^*$, design matrix $\bxl^*$ 
and likelihood function raised to the power of $1/\delta$, is given by 
$$
m_\ell^N(\by^* | \bxl^*, \delta) = c_\ell \, \pi^{\frac{1}{2}(d_\ell-n^*)} | {\bxl^T}^* \bxl^* |^{-\frac{1}{2}} \Gamma \left( \frac{n^*-d_\ell}{2} \right) {RSS_\ell^*}^{-\frac{n^*-d_\ell}{2}}, 
$$
where $RSS_\ell^*$ is the residual sum of squares given by $RSS_\ell^* = {\by^*}^T \mt{P}_\ell^* \, \by^*$. Similarly, in the rest of the paper we denote by $RSS_\ell = {\by}^T \mt{P}_\ell \, \by$.

The J-PEP prior for the parameters of model $M_\ell$ is given by 
\begin{eqnarray*}
\pi_\ell^{J-PEP} ( \bbl, \varl | \bxl^*, \delta) 
&=&  \int \pi_\ell^{ N } ( \dn{ \beta }_\ell, \sigma_\ell^2 | \dn{ y }^*
;  \mt{ X }_\ell^*, \delta ) m_0^N( \by^* | \bxo^*, \delta) d \by^*  \\
&=&  \int f( \by^* | \bbl, \varl ,  M_\ell; \bxl^*,\delta) \pi^N( \bbl, \varl | \bxl^* ) 
\frac{m_0^N( \by^* | \bxo^*, \delta)}{m_\ell^N( \by^* | \bxl^*, \delta)} d \by^*  \\
&=&  \int \int \left[ 
     \int \frac{f( \by^* | \bbl, \varl ,  M_\ell; \bxl^*,\delta) f( \by^* | \bbo, \varo , M_0;\bx_0^*, \delta)  \pi^N( \bbl, \varl | \bxl^* ) 
     }{m_\ell^N( \by^* | \bxl^*, \delta)} d \by^* \right] 
     \\ 
     && \hspace{21.5em}\times \pi_0^N( \bbo, \varo |\bxo^* ) d\bbo d\varo \\ 
&=&  \int \int \pi_\ell^{J-PEP} ( \bbl, \varl | \bbo, \varo; \bxl^*, \delta )  \pi_0^N( \bbo, \varo|\bxo^*) d\bbo d\varo
\end{eqnarray*}
with the conditional J-PEP prior given by 

\begin{multline}
\pi_\ell^{J-PEP} ( \bbl, \varl | \bbo, \varo; \bxl^*, \delta ) 
=  \int \frac{ f_{N_{\nstar}}( \by^* ;  \bxl \bbl, \delta \varl \identst) 
                 f_{N_{\nstar}}( \by^* ;  \bxo \bbo, \delta \varo \identst) c_\ell/\varl }
               {c_\ell \, \pi^{\frac{1}{2}(d_\ell-n^*)} | {\bxl^*}^T \bxl^* |^{-\frac{1}{2}} \Gamma \left( \frac{n^*-d_\ell}{2} \right)
               {RSS_\ell^*}^{-\frac{n^*-d_\ell}{2}}} d \by^*     \\ 
=  \frac{\pi^{-\frac{1}{2}(d_\ell-n^*)}  }{\varl \Gamma \left( \frac{n^*-d_\ell}{2} \right)} | {\bxl^*}^T \bxl^* |^{ \frac{1}{2}} 
							\times \int {RSS_\ell^*}^{\frac{n^*-d_\ell}{2}}
                                f_{N_{\nstar}}( \by^* ;  \bxl \bbl, \delta \varl \identst) 
                                f_{N_{\nstar}}( \by^* ;  \bxl \bboup, \delta \varo \identst) d \by^*     
                                \label{condprior1}
\end{multline}
where $\bboup = ( \bbo^T, \dn{0}_{ d_\ell-d_0 }^T )^T$ and $\dn{0}_{k}$ being a vector of zeros of length $k$. 
The product of the two normal densities involved in the integrand is given by 
$$
f_{N_{\nstar}}( \by^* ;  \bxl \bbl, \delta \varl \identst) 
f_{N_{\nstar}}( \by^* ;  \bxl \bboup, \delta \varo \identst) = \mbox{\hspace{9cm}}
$$
\vspace{-2em}
\begin{eqnarray} 
\hspace{5cm}&=& 
(2\pi)^{-\frac{\nstar-d_\ell}{2}} 
\left[ \delta ( \varo + \varl ) \right]^{-\frac{\nstar-d_\ell}{2}}
|{\bxl^*}^T\bxl^*|^{-\frac{1}{2}} \nonumber
f_{N_{\nstar}}\left(  \by^* ; \mt{E}^{-1} \mt{D}, \mt{E}^{-1} \right) \\ && \times
f_{N_{d_\ell}}\left(  \bbl ;  \bboup,  \delta(\varl + \varo) \invxtxstarl \right) 
\label{cprior_point1}
\end{eqnarray} 
with 
\begin{equation}
\mt{E} = \left( \frac{\varl + \varo}{\delta \varo \varl }\right) \identst \mbox{~and~} 
\mt{D} = \frac{1}{\delta \varo} \bxl^* \bboup + \frac{1}{\delta \varl} \bxl^* \bbl 
= \frac{1}{\delta}  \bxl^* \left( \frac{\varl}{\varl + \varo}  \bboup + \frac{\varo}{\varl + \varo} \bbl \right)~. 
\label{ED}
\end{equation}
Note that (\ref{cprior_point1}) was obtained using the property 
\begin{eqnarray} 
f_{N_n}( \by ;  \mt{M} \dn{\xi}_1, \mt{A}_1) 
f_{N_n}( \by ;  \mt{M} \dn{\xi}_2, \mt{A}_2) 
&=& 
(2\pi)^{-\frac{n-p}{2}} 
|\mt{A}_1+\mt{A}_2|^{-\frac{1}{2}} | \mt{M}^T( \mt{A}_1+\mt{A}_2 )^{-1}\mt{M} |^{-\frac{1}{2}} \nonumber \\ && \times
f_{N_{n}}\left(  \by ; \mt{E}_1^{-1} \mt{D}_1, \mt{E}_1^{-1} \right)
f_{N_{n}}\left(  \dn{\xi}_1 ;  \dn{\xi}_2,  \mt{A}_1+\mt{A}_2 \right) 
\label{cprior_gen}  
\end{eqnarray} 
with
$$
\mt{E}_1 = \mt{A}_1^{-1} + \mt{A}_2^{-1} \mbox{~and~} 
\mt{D}_1 = \mt{A}_1^{-1}\mt{M} \dn{\xi}_1 + \mt{A}_2^{-1} \mt{M} \dn{\xi}_2~. 
$$
In (\ref{cprior_gen}), $\mt{M}$ is a $n \times p$ matrix of rank $p$ ($p \leq n$), $\dn{\xi}_1$ and $\dn{\xi}_1$ are vectors of length $p$ and $\mt{A}_1$ and $\mt{A}_2$ are positive definite matrices of dimension $n \times n$.
Expression (\ref{cprior_gen}) can be easily obtained using the identity:
$$
(\by-\mt{M}\dn{\xi}_1)^T\mt{A}_1^{-1}(\by-\mt{M}\dn{\xi}_1)+(\by-\mt{M}\dn{\xi}_2)^T\mt{A}_2^{-1}(\by-\mt{M}\dn{\xi}_2) = \hspace{7cm}
$$
\vspace{-2.5em}
\begin{eqnarray*}
\hspace{3cm}
&=&\by^T\mt{E}\by-2\by^T(\mt{A}_1^{-1}\mt{M}\dn{\xi}_1+\mt{A}_2^{-1}\mt{M}\dn{\xi}_1)+\dn{\xi}_1^T+\mt{M}^T\mt{A}_1^{-1}\mt{M}\dn{\xi}_1+\dn{\xi}_2^T+\mt{M}^T\mt{A}_2^{-1}\mt{M}\dn{\xi}_2 \nonumber \\
&=&[\mt{C}^T \by -\mt{C}^{-1}\mt{D}]^T [\mt{C}^T \by -\mt{C}^{-1}\mt{D}]+(\dn{\xi}_2-\dn{\xi}_1)^T \mt{M}^T(\mt{A}_1+\mt{A}_2)^{-1} \mt{M} (\dn{\xi}_2-\dn{\xi}_1), \nonumber
\end{eqnarray*}
with $\mt{C}$ being a $n \times n$ lower triangular matrix (the Cholesky decomposition) with non zero elements in the diagonal such that $\mt{E}_1=\mt{C}\mt{C}^T$.

Replacing (\ref{cprior_point1}) in (\ref{condprior1}), we obtain 
\begin{eqnarray}
\pi_\ell^{J-PEP} ( \bbl, \varl | \bbo, \varo; \bxl^*, \delta ) 
&=&  \frac{\pi^{-\frac{1}{2}(d_\ell-n^*)}  }{\varl \Gamma \left( \frac{n^*-d_\ell}{2} \right)}  |{\bxl^*}^T \bxl^* |^{ \frac{1}{2}} 
							(2\pi)^{-\frac{\nstar-d_\ell}{2}} \left[ \delta ( \varo + \varl ) \right]^{-\frac{\nstar-d_\ell}{2}}
							|{\bxl^*}^T\bxl^*|^{-\frac{1}{2}} 
							\nonumber \\ && 
							\times  f_{N_{\nstar}}\left(  \bbl ;  \bboup,  \delta(\varl + \varo) \invxtxstarl \right) \nonumber \\ &&
							\times \int \big({\by^*}^T \mt{P}_\ell^* \, \by^*\big)^{\frac{n^*-d_\ell}{2}}
							           f_{N_{\nstar}}\left(  \by^* ; \mt{E}^{-1} \mt{D}, \mt{E}^{-1} \right) d \by^* ,    
							           \label{condprior2}
\end{eqnarray}
with $\mt{E}$ and $\mt{D}$ given in (\ref{ED}).

We set 
$$
\dn{z}= \mt{E}^{\frac{1}{2}} \left( \by^* - \mt{E}^{-1} \mt{D}\right) = \zeta^{\frac{1}{2}}( \by^* - \bxl^* \mt{\Gamma} )
$$ 
where $\zeta = \left( \frac{\varl + \varo}{\delta \varo \varl }\right)$  
and
$\mt{\Gamma} = (\zeta \delta)^{-1}  \left( \frac{\varl}{\varl + \varo}  \bboup + \frac{\varo}{\varl + \varo} \bbl \right)$. 
Therefore we have 
$\by^*=\zeta^{-1/2}\dn{z}+\bxl^* \mt{\Gamma}$, $d\by^*=\zeta^{-\nstar/2} d\dn{z}$ and 
$$
f_{N_{\nstar}}\left(  \by^* ; \mt{E}^{-1} \mt{D}, \mt{E}^{-1} \right) d \by^* = 
f_{N_{\nstar}}\left(  \dn{z} ; \dn{0}_{\nstar}, \identst \right) d \dn{z}
$$
since the term $\zeta^{-\nstar/2}$, coming from the Jacobian of the transformation, cancels out 
with the determinant of the variance, that is $|\mt{E}|^{1/2} = \zeta^{n^*/2}$.
Moreover, 
\begin{eqnarray}
{\by^*}^T \mt{P}_\ell^* \, \by^*
&=& (\zeta^{-\frac{1}{2}} \dn{z} + \bxl^* \mt{\Gamma})^T \mt{P}_\ell^* \, (\zeta^{-\frac{1}{2}} \dn{z} + \bxl^* \mt{\Gamma})\big) \nonumber \\ 
&=& \zeta^{-1} \dn{z}^T  \mt{P}_\ell^*\dn{z} 
+ \zeta^{-\frac{1}{2}} \dn{z}^T  \mt{P}_\ell^*\bxl^* \mt{\Gamma}
+ \mt{\Gamma}^T  {\bxl^*}^T \mt{P}_\ell^*\zeta^{-\frac{1}{2}} \dn{z} 
+ \mt{\Gamma}^T {\bxl^*}^T \mt{P}_\ell^*\bxl^* \mt{\Gamma}\nonumber \\ 
&=& \zeta^{-1} \dn{z}^T  \mt{P}_\ell^*\dn{z} 
\end{eqnarray}
since ${\bxl^*}^T \mt{P}_\ell^*=\mt{P}_\ell^*{\bxl^*} =\dn{0}$. 

Returning back to (\ref{condprior2}) we obtain 
$$
\pi_\ell^{J-PEP} ( \bbl, \varl | \bbo, \varo; \bxl^*, \delta ) = \hspace{12cm}
$$
\vspace{-1em}
\begin{eqnarray}
&=&  2^{-\frac{\nstar-d_\ell}{2}} \left[ \varl \Gamma \left( \frac{n^*-d_\ell}{2} \right) \right]^{-1} 
              \left[ \delta ( \varo + \varl ) \right]^{-\frac{\nstar-d_\ell}{2}} 
							       f_{N_{\nstar}}\left(  \bbl ;  \bboup,  \delta(\varl + \varo) \invxtxstarl \right) \nonumber \\ &&
							\hspace{7.9cm} \times ~ \zeta^{-\frac{n^*-d_\ell}{2}} \int \big( \dn{z}^T \mt{P}_\ell^* \dn{z} \big)^{\frac{n^*-d_\ell}{2}}
							           f_{N_{\nstar}}\left(  \dn{z} ; \dn{0}_{\nstar}, \identst \right) d \dn{z}      \nonumber \\   
&=&  2^{-\frac{\nstar-d_\ell}{2}} \left[  \Gamma \left( \frac{n^*-d_\ell}{2} \right) \right]^{-1} 
              \left[ \delta ( \varo + \varl ) \right]^{-\frac{\nstar-d_\ell}{2}} 
							\delta^{ \frac{\nstar-d_\ell}{2} } (\varo)^{ \frac{\nstar-d_\ell}{2} } (\varl)^{ \frac{\nstar-d_\ell}{2} -1 } 
							(\varo+\varl)^{ -\frac{\nstar-d_\ell}{2} }
							\nonumber \\ && 
							\hspace{6cm} \times  f_{N_{\nstar}}\left(  \bbl ;  \bboup,  \delta(\varl + \varo) \invxtxstarl \right)  
							  E\left[(\dn{z}^T\mt{P}_\ell^*\dn{z})^{\frac{\nstar-d_\ell}{2}} \right]  \nonumber \\   
&=&  2^{-\frac{\nstar-d_\ell}{2}} \left[  \Gamma \left( \frac{n^*-d_\ell}{2} \right) \right]^{-1} 
							(\varo)^{ \frac{\nstar-d_\ell}{2} } (\varl)^{ \frac{\nstar-d_\ell}{2} -1 } 
							(\varo+\varl)^{ -(\nstar-d_\ell) }
							\nonumber \\ && 
							\hspace{4.9cm} \times  f_{N_{\nstar}}\left(  \bbl ;  \bboup,  \delta(\varl + \varo) \invxtxstarl \right) \nonumber   
							 2^{\frac{\nstar-d_\ell}{2}} \frac{\Gamma( \frac{\nstar-d_\ell}{2}+\frac{\nstar-d_\ell}{2} )}{\Gamma(\frac{\nstar-d_\ell}{2})},  
							\label{condprior3}
\end{eqnarray}
since 
$$
E\left[(\dn{x}^TK\dn{x})^{h} \right] = 2^h \frac{\Gamma(h+r/2)}{r/2}
$$
where $h >0$, $K$ is a $n \times n$ symmetric and idempotent matrix of rank $r$, $\dn{x} \sim N_n(\dn{0}_n, \ident)$ and therefore $\dn{x}^TK\dn{x} \sim \chi^2_{r}$. 

Thus (\ref{condprior3}) becomes 
\begin{eqnarray}
\pi_\ell^{J-PEP} ( \bbl, \varl | \bbo, \varo; \bxl^*, \delta ) 
&=&  \frac{\Gamma( \nstar-d_\ell )}{\Gamma(\frac{\nstar-d_\ell}{2})^2}  
							(\varo)^{ -\frac{\nstar-d_\ell}{2} } (\varl)^{ \frac{\nstar-d_\ell}{2} -1 } 
							\left( 1 + \frac{\varl}{\varo} \right)^{ -(\nstar-d_\ell) }
							\nonumber \\ && 
							\times  f_{N_{d_\ell}}\left(  \bbl ;  \bboup,  \delta(\varl + \varo) \invxtxstarl \right)~.  
\label{finalcondprior}
\end{eqnarray}

\section{The J-PEP Bayes factor} 


The Bayes factor of any model $M_\ell$ ($\ell \neq 0$) versus the reference model $M_0$, under the J-PEP prior approach, is given by 
$$
BF_{\ell \,0}^{J-PEP} = \frac
{  \int f_{N_n}( \by ; \bxl \bbl, \varl \ident ) \pi_\ell^{J-PEP} ( \bbl, \varl | \bxl^*, \delta) d\bbl d\varl }
{  \int f_{N_n}( \by ; \bxo \bbo, \varo \ident ) \pi_0^{N}( \bbo, \varo | \mt{X}_0^*)  d\bbo d\varo }
$$
with the denominator given by 
$$
m_0^N(\by| \mt{X}_0) = c_0 \pi^{\frac{1}{2}(d_0-n)} |\bxo^T \bxo|^{-\frac{1}{2}} \Gamma \left( \frac{n-d_0}{2} \right) RSS_0^{-\frac{n-d_0}{2}}~. 
$$

Using (\ref{finalcondprior}), the numerator is given by 
$$
m^{ J-PEP }_\ell ( \dn{ y } | \mt{ X }_\ell \, , \mt{ X }_\ell^* \, , \delta ) = \hspace{13cm}
$$
\vspace{-1.5em}
\begin{eqnarray*}
&=& \int\int\int\int f_{N_n}( \by ; \bxl \bbl, \varl \ident ) \pi_\ell^{J-PEP} ( \bbl, \varl | \bbo, \varo; \bxl^*, \delta ) \pi_0^N( \bbo, \varo | \mt{X}_0^*) d\bbl d\varl d\bbo d\varo \\ 
&=& \int\int\int\int \frac{c_0}{\varo} C_\ell
f_{N_n}( \by ; \bxl \bbl, \varl \ident ) 
f_{N_{d_\ell}} \left( \bbl ; \bboup, \delta (\varl + \varo) \invxtxstarl \right) d\bbl d\varl d\bbo d\varo, 
\end{eqnarray*}
with 
\be
C_\ell 
= (\varo)^{-\frac{\nstar-d_\ell}{2}} (\varl)^{\frac{\nstar-d_\ell}{2}-1} \left( 1 + \frac{\varl}{\varo} \right)^{-(\nstar-d_\ell)} 
\frac{ \Gamma\left( n^*-d_\ell \right) }{ \Gamma\left( \frac{n^*-d_\ell}{2} \right)^2 }.
\label{c_ell}
\ee

Integrating out $\bbl$, we obtain 
\begin{eqnarray*}
m^{ J-PEP }_\ell ( \dn{ y } | \mt{ X }_\ell \, , \mt{ X }_\ell^* \, , \delta ) 
&=& \int\int\int \frac{c_0}{\varo} C_\ell \left[f_{N_n}( \by ; \bxl \bboup, \doS ) \right] d\bbo  d\varl  d\varo, 
\end{eqnarray*}
with 
$$
\doS = 
\varl \ident + \delta ( \varl + \varo ) \bxl \invxtxstarl \bxl^T ~. 
$$
The above expression was obtained using the following formula:
$$
\int f_{N_n}( \by ; \mt{M} \dn{\xi}_1, \mt{A}_1 ) f_{N_{p}} \left( \dn{\xi}_1 ; \dn{\xi}_2, \mt{A}_3 \right) d\dn{\xi}_1 
= 
f_{N_n}( \by ; \mt{M} \dn{\xi}_2, \mt{A}_1 + \mt{M} \mt{A}_3 \mt{M}^T )~,  
$$
with $\mt{M}$ being a $n \times p$ matrix of rank $p$ ($p \leq n$), $\dn{\xi}_1$ and $\dn{\xi}_2$ being vectors of length $p$ and $\mt{A}_1$ and $\mt{A}_3$ being positive definite matrices of dimensions $n \times n$ and $p \times p$ respectively.

Moreover, 
\begin{eqnarray*}
m^{ J-PEP }_\ell ( \dn{ y } | \mt{ X }_\ell \, , \mt{ X }_\ell^* \, , \delta )\hspace{-0.7em}
&=& \hspace{-0.7em} \int\int\int \frac{c_0}{\varo}C_\ell  \left[f_{N_n}( \by ; \bxl \bboup, \doS ) \right] d\bbo  d\varl  d\varo \\
&=& \hspace{-0.7em} \int\int\int \frac{c_0}{\varo}C_\ell  \left[f_{N_n}( \by ; \bxo \bbo, \doS ) \right] d\bbo  d\varl  d\varo \\ 
&=& \hspace{-0.7em} \int\int \frac{c_0}{\varo}C_\ell  \left[
(2\pi)^{- \frac{n-d_0}{2}} 
|\doS|^{-\frac{1}{2}}
|\bxo^T \doS^{-1} \bxo|^{-\frac{1}{2}}
\exp \left\{ - \frac{1}{2} \by^T A_{\Sigma} \by \right\} \right] d\varl  d\varo, 
\end{eqnarray*}
where 
$$
A_{\Sigma} = \doS^{-1} -\doS^{-1} \bxo 
\left[ \bxo^T \doS^{-1} \bxo\right]^{-1} \bxo^T \doS^{-1},
$$
since 
\begin{eqnarray*}
\int f_{N_n}( \by ; \mt{M} \dn{\xi}_1, \mt{A}_1)  d\dn{\xi}_1
&=&  
(2\pi)^{ - \frac{n-p}{2} } | \mt{A}_1 |^{-\frac{1}{2}} |\mt{M}^T \mt{A}_1^{-1} \mt{M}|^{-\frac{1}{2}} \\ &&\times
\exp \left\{ -\frac{1}{2} \by^T \left[ \mt{A}_1^{-1} -  \mt{A}_1^{-1} \mt{M} \big( \mt{M}^T \mt{A}_1^{-1} \mt{M} \big)^{-1} \mt{M}^T \mt{A}_1^{-1} \right] \by\right\} 
\end{eqnarray*}
with $\mt{M}$ being a $n \times p$ matrix of rank $p$ ($p \leq n$), $\dn{\xi}_1$ being a vector of length $p$ and $\mt{A}_1$ being a positive definite matrix of dimension $n \times n$. 

Substituting expression (\ref{c_ell}), we obtain 
\begin{eqnarray}
m^{ J-PEP }_\ell ( \dn{ y } | \mt{ X }_\ell \, , \mt{ X }_\ell^* \, , \delta ) 
&=& \int\int \frac{c_0}{\varo}
(\varo)^{-\frac{\nstar-d_\ell}{2}} (\varl)^{\frac{\nstar-d_\ell}{2}-1} \left( 1 + \frac{\varl}{\varo} \right)^{-(\nstar-d_\ell)} 
\frac{ \Gamma\left( n^*-d_\ell \right) }{ \Gamma\left( \frac{n^*-d_\ell}{2} \right)^2 } \nonumber\\ 
&&\times 
\left[
(2\pi)^{- \frac{n-d_0}{2}} 
|\doS|^{-\frac{1}{2}}
|\bxo^T \doS^{-1} \bxo|^{-\frac{1}{2}}  
\exp \left\{ - \frac{1}{2} \by^T A_{\Sigma} \by \right\} \right] d\varl  d\varo \nonumber\\ 
&=& c_0 (2\pi)^{- \frac{n-d_0}{2}} \frac{ \Gamma\left( n^*-d_\ell \right) }{ \Gamma\left( \frac{n^*-d_\ell}{2} \right)^2 } 
 \int\int (\varo)^{-2}
\left( \frac{ \varl }{ \varo } \right)^{\frac{\nstar-d_\ell}{2}-1} \left( 1 + \frac{\varl}{\varo} \right)^{-(\nstar-d_\ell)} \nonumber\\
&& \times |\doS|^{-\frac{1}{2}}|\bxo^T \doS^{-1} \bxo|^{-\frac{1}{2}}  \exp \left\{ - \frac{1}{2} \by^T A_{\Sigma} \by \right\}  d\varl  d\varo. \label{marginal1} 
\end{eqnarray}

We now set 
$$
r= \sqrt{ \varo + \varl } \mbox{~and~} \phi = \arctan \left( \sqrt{ \frac{\varl}{\varo} } \right)
$$
for $r \in [0, +\infty)$  and $\phi \in [0, \pi/2]$. 
The inverse transformations are given by 
\be
\varo = r^2 \cos^2 \phi \mbox{~and~} \varl = r^2 \sin^2 \phi
\label{transf}
\ee
while the Jacobian is 
\begin{eqnarray}
J(r,\phi) 
&= &
\left| 
\begin{array}{cc} 
\frac{ \partial \varo }{ \partial r } & \frac{ \partial \varo }{ \partial \phi }\\ 
\frac{ \partial \varl }{ \partial r } & \frac{ \partial \varl }{ \partial \phi }\\  
\end{array}
\right|
= 
\left| 
\begin{array}{cc} 
\frac{ \partial (r^2 \cos^2 \phi) }{ \partial r } & \frac{ (\partial r^2 \cos^2 \phi) }{ \partial \phi }\\ 
\frac{ \partial (r^2 \sin^2 \phi) }{ \partial r } & \frac{ (\partial r^2 \sin^2 \phi) }{ \partial \phi }\\  
\end{array}
\right|
= 
\left| 
\begin{array}{cc} 
2r \cos^2 \phi & -2 r^2 \cos \phi \sin \phi \\ 
2r \sin^2 \phi &  2 r^2 \sin \phi \cos \phi  \\  
\end{array}
\right| \nonumber \\ 
&=&  4r^3 \sin \phi \cos \phi (  \cos^2\phi + \sin^2 \phi) = 4r^3 \sin \phi \cos \phi ~. 
\label{Jacobian} 
\end{eqnarray}
Then, the matrix $\doS$ becomes equal to  
\begin{eqnarray}
\doS 
&=& \varl \ident + \delta ( \varl + \varo ) \bxl \invxtxstarl \bxl^T  
 =  r^2 \sin^2 \phi ~\ident + r^2 \delta \bxl \invxtxstarl \bxl^T  
 = r^2 B(\phi)
\label{marginal1_p1} 
\end{eqnarray}
with $B(\phi)$ being a $n\times n$ matrix given by 
\be
B(\phi) = \sin^2 \phi ~\ident +  \delta \bxl \invxtxstarl \bxl^T ~
\label{bphi}
\ee
while $A_\Sigma$ can be rewritten as 
\begin{eqnarray}
A_\Sigma 
&=& \doS^{-1} -\doS^{-1} \bxo \left[ \bxo^T \doS^{-1} \bxo\right]^{-1} \bxo^T \doS^{-1}  \nonumber \\ 
&=&  r^{-2} B^{-1}(\phi) -r^{-2} B^{-1}(\phi) \bxo \left[ \bxo^T r^{-2} B^{-1}(\phi) \bxo\right]^{-1} \bxo^T r^{-2} B^{-1}(\phi) \nonumber\\ 
&=&  r^{-2} \left[ B^{-1}(\phi) -  B^{-1}(\phi) \bxo A^{-1}(\phi) \bxo^T B^{-1}(\phi) \right] \nonumber
\end{eqnarray}
with 
\be
A(\phi) = \bxo^T B^{-1}(\phi) \bxo 
\label{aphi}
\ee
being a $d_0 \times d_0$ matrix. 
Moreover, we have that 
\be
\by^T A_{\Sigma} \by = r^{-2} D(\phi)
\label{marginal1_p2}
\ee
with
\be
D(\phi) = \by^T \left[ B^{-1}(\phi) -  B^{-1}(\phi) \bxo A^{-1}(\phi) \bxo^T B^{-1}(\phi) \right] \by 
\label{ephi}
\ee
being a scalar. 
Finally, the first three terms in the integrand of (\ref{marginal1}) can be written as
$$
(\varo)^{-2}
\left( \frac{ \varl }{ \varo } \right)^{\frac{\nstar-d_\ell}{2}-1} \left( 1 + \frac{\varl}{\varo} \right)^{-(\nstar-d_\ell)} = 
\hspace{24em}
$$
\vspace{-1.5em} 
\begin{eqnarray}
\hspace{10em}
&=& (r^2 \cos^2 \phi )^{-2}
\left( \frac{ \sin^2 \phi }{ \cos^2 \phi } \right)^{\frac{\nstar-d_\ell}{2}-1} \left( \frac{r^2 \cos^2 \phi + r^2 \sin^2 \phi}{r^2 \cos^2 \phi} \right)^{-(\nstar-d_\ell)} \nonumber\\ 
&=& 
(r^2 \cos^2 \phi )^{-2}
\left( \frac{ \sin^2 \phi }{ \cos^2 \phi } \right)^{\frac{\nstar-d_\ell}{2}-1} 
(\cos^2 \phi)^{\nstar-d_\ell} \nonumber\\ 
&=&  r^{-4} (\sin\phi \cos \phi)^{\nstar-d_\ell-2}. 
\label{marginal1_p3}
\end{eqnarray}

Using the transformation (\ref{transf}) and the corresponding Jacobian given by (\ref{Jacobian}), 
as well as expressions (\ref{marginal1_p1}), (\ref{marginal1_p2}) and (\ref{marginal1_p3}), 
the marginal likelihood (\ref{marginal1}) now becomes 
$$
m^{ J-PEP }_\ell ( \dn{ y } | \mt{ X }_\ell \, , \mt{ X }_\ell^* \, , \delta ) =
\hspace{15cm}
$$
\vspace{-2em} 
\begin{eqnarray}
\hspace{4em}&=& c_0 (2\pi)^{- \frac{n-d_0}{2}} 
\frac{ \Gamma\left( n^*-d_\ell \right) }{ \Gamma\left( \frac{n^*-d_\ell}{2} \right)^2 }  
\int \limits _{0}^{\pi/2} \ \int \limits _{0}^{\infty} \ 
\frac{ r^{-4} (\sin\phi \cos \phi)^{\nstar-d_\ell-2}}
     { |r^2 B(\phi)|^{\frac{1}{2}} |r^{-2} \bxo^T B^{-1}(\phi) \bxo|^{\frac{1}{2}} }  \nonumber\\  
&& \times  \exp \left\{ - \frac{1}{2} r^{-2}D(\phi) \right\} 4r^3 \sin \phi \cos \phi ~  drd\phi  \nonumber \\
&=& 4 c_0 (2\pi)^{- \frac{n-d_0}{2}} 
\int \limits _{0}^{\pi/2} \ \frac{(\sin\phi \cos \phi)^{\nstar-d_\ell-1}}
                                 {|B(\phi)|^{\frac{1}{2}} | \bxo^T B^{-1}(\phi) \bxo|^{\frac{1}{2}} } 
                                 \int \limits _{0}^{\infty} \ r^{-n+d_0-1} 
                                 \exp \left\{ - \frac{1}{2} r^{-2}D(\phi) \right\}  ~  dr d\phi . \nonumber\\ 
\end{eqnarray}

We now set $w=1/r$ ($\Leftrightarrow r=w^{-1}$ and $dr = (-1)w^{-2}dw$), resulting in 
\begin{eqnarray}
m^{ J-PEP }_\ell ( \dn{ y } | \mt{ X }_\ell \, , \mt{ X }_\ell^* \, , \delta ) 
&=& 4 c_0 (2\pi)^{- \frac{n-d_0}{2}} 
\frac{ \Gamma\left( n^*-d_\ell \right) }{ \Gamma\left( \frac{n^*-d_\ell}{2} \right)^2 }  \nonumber \\ && \times
\int \limits _{0}^{\pi/2} \ \frac{ (\sin\phi \cos \phi)^{\nstar-d_\ell-1} }
                                 { |B(\phi)|^{\frac{1}{2}} | A(\phi)|^{\frac{1}{2}} }   
                                 \int \limits _{0}^{\infty} \ w^{n-d_0+1} 
                                 \exp \left\{ - \frac{1}{2} w^2 D(\phi) \right\}  ~  w^{-2}dw d\phi  \nonumber\\ 
&=& 4 c_0 (2\pi)^{- \frac{n-d_0}{2}} 
\frac{ \Gamma\left( n^*-d_\ell \right) }{ \Gamma\left( \frac{n^*-d_\ell}{2} \right)^2 }  \nonumber \\ && \times
\int \limits _{0}^{\pi/2} \ \frac{ (\sin\phi \cos \phi)^{\nstar-d_\ell-1} }
                                 { |B(\phi)|^{\frac{1}{2}} | A(\phi)|^{\frac{1}{2}} D(\phi)}     
                                 \int \limits _{0}^{\infty} \ w^{n-d_0-2} \frac{w}{D(\phi)^{-1}}
                                 \exp \left\{ -\frac{w^2}{2D(\phi)^{-1}} \right\}  dw d\phi  \nonumber
\end{eqnarray}                                 
\begin{eqnarray}                                 
&=& 4 c_0 (2\pi)^{- \frac{n-d_0}{2}} 
\frac{ \Gamma\left( n^*-d_\ell \right) }{ \Gamma\left( \frac{n^*-d_\ell}{2} \right)^2 }   
\int \limits _{0}^{\pi/2} \frac{ (\sin\phi \cos \phi)^{\nstar-d_\ell-1} }
                                 { |B(\phi)|^{\frac{1}{2}} | A(\phi)|^{\frac{1}{2}} D(\phi)}  
                                 \int \limits _{0}^{\infty} \ w^{n-d_0-2} f_R( w; D(\phi)^{-1} ) ~dw d\phi \nonumber\\
&=& 4 c_0 (2\pi)^{- \frac{n-d_0}{2}} 
\frac{ \Gamma\left( n^*-d_\ell \right) }{ \Gamma\left( \frac{n^*-d_\ell}{2} \right)^2 }  
\int \limits _{0}^{\pi/2} \ \frac{ (\sin\phi \cos \phi)^{\nstar-d_\ell-1} }
                                 { |B(\phi)|^{\frac{1}{2}} | A(\phi)|^{\frac{1}{2}} D(\phi)}  
                                 E_R (  w^{n-d_0-2} ; D(\phi)^{-1} )  d\phi, \nonumber
\end{eqnarray}
where $f_R( w; s^2 )$ is the density function of the Rayleigh distribution with scale parameter $s^2$
(which here is equal to $D(\phi)^{-1}$) and variance $s^2(4-\pi)/2$. 
Moreover, by $E_R (  w^k ; s^2 )$  we denote the corresponding $k^{th}$ moment about zero which is given by $s^k 2^{k/2} \Gamma(1+k/2)$. 
Therefore we have:  
$$
m^{ J-PEP }_\ell ( \dn{ y } | \mt{ X }_\ell \, , \mt{ X }_\ell^* \, , \delta ) = \hspace{15cm}
$$
\vspace{-3em} 
\begin{eqnarray}
&=& 4 c_0 (2\pi)^{- \frac{n-d_0}{2}} 
\frac{ \Gamma\left( n^*-d_\ell \right) }{ \Gamma\left( \frac{n^*-d_\ell}{2} \right)^2 }  
\int \limits _{0}^{\pi/2} \ \frac{ (\sin\phi \cos \phi)^{\nstar-d_\ell-1} 2^{\frac{n-d_0-2}{2}} \Gamma\left(1+\frac{n-d_0-2}{2}\right) }
                                 { |B(\phi)|^{\frac{1}{2}} | A(\phi)|^{\frac{1}{2}} [D(\phi)]^{1+\frac{n-d_0-2}{2}}} 
                                 d\phi \nonumber \\ 
&=& 4 c_0 (2\pi)^{- \frac{n-d_0}{2}} 
\frac{ \Gamma\left( n^*-d_\ell \right) }{ \Gamma\left( \frac{n^*-d_\ell}{2} \right)^2 }  
2^{\frac{n-d_0}{2}-1} \Gamma\left(\frac{n-d_0}{2}\right)
\int \limits _{0}^{\pi/2} \ \frac{ (\sin\phi \cos \phi)^{\nstar-d_\ell-1}  }
                                 { |B(\phi)|^{\frac{1}{2}} | A(\phi)|^{\frac{1}{2}} [D(\phi)]^{\frac{n-d_0}{2}}} 
                                 d\phi \nonumber\\
&=& 2c_0 \pi^{- \frac{n-d_0}{2}} 
\frac{ \Gamma\left( n^*-d_\ell \right) \Gamma\left(\frac{n-d_0}{2}\right)}{ \Gamma\left( \frac{n^*-d_\ell}{2} \right)^2 }  
\int \limits _{0}^{\pi/2} \ \frac{ (\sin\phi \cos \phi)^{\nstar-d_\ell-1}  }
                                 { |B(\phi)|^{\frac{1}{2}} | A(\phi)|^{\frac{1}{2}} [D(\phi)]^{\frac{n-d_0}{2}}} 
                                 d\phi. \nonumber                               
\end{eqnarray}

Hence the Bayes factor of model $M_\ell$ ($\ell \neq 0$) versus the reference model $M_0$, under the J-PEP prior approach, is given by 
\begin{eqnarray}
BF_{\ell \,0}^{J-PEP}
&=& \frac{ 2c_0 \pi^{- \frac{n-d_0}{2}} 
\frac{ \Gamma\left( n^*-d_\ell \right) \Gamma\left(\frac{n-d_0}{2}\right)}{ \Gamma\left( \frac{n^*-d_\ell}{2} \right)^2 }  }
{ c_0 \pi^{\frac{1}{2}(d_0-n)} |\bxo^T \bxo|^{-\frac{1}{2}} \Gamma \left( \frac{n-d_0}{2} \right) RSS_0^{-\frac{n-d_0}{2}}}
\int \limits _{0}^{\pi/2} \ \frac{ (\sin\phi \cos \phi)^{\nstar-d_\ell-1}  }
                                 { |B(\phi)|^{\frac{1}{2}} | A(\phi)|^{\frac{1}{2}} [D(\phi)]^{\frac{n-d_0}{2}}} 
                                 d\phi. \nonumber \\
&=& 2 ~
\frac{ \Gamma\left( n^*-d_\ell \right) }{ \Gamma\left( \frac{n^*-d_\ell}{2} \right)^2 } 
|\bxo^T \bxo|^{ \frac{1}{2} } RSS_0^{\frac{n-d_0}{2}}
\int \limits _{0}^{\pi/2} \ \frac{ (\sin\phi \cos \phi)^{\nstar-d_\ell-1}  }
                                 { |B(\phi)|^{\frac{1}{2}} | A(\phi)|^{\frac{1}{2}} [D(\phi)]^{\frac{n-d_0}{2}}} 
                                 d\phi.  \label{bf1}
\end{eqnarray}

Under the J-PEP approach we set $\xtxstarl = \xtxl$, $n^*=n$ and $\delta=n$ and thus 
 
$$
B(\phi) 
=  \sin^2 \phi ~\ident +  \delta \bxl \invxtxl \bxl^T
=  \sin^2 \phi ~\ident +  \delta \mt{H}_\ell~.
$$
Moreover, 
\begin{eqnarray} 
B^{-1}(\phi) 
&=&  [\sin^2 \phi ~\ident +  \delta \mt{H}_\ell]^{-1} 
 =   \frac{1}{\sin^2 \phi} \left[ \ident +  \frac{\delta}{\sin^2 \phi} \bxl \invxtxl \bxl^T \right]^{-1}  \nonumber\\
&=&  \frac{1}{\sin^2 \phi} 
\left[ \ident^{-1} - \ident^{-1} \frac{\delta}{\sin^2 \phi} \bxl \left( \left[\invxtxl\right]^{-1} +\frac{\delta}{\sin^2 \phi} \bxl^T \bxl \right)^{-1} \bxl^T \ident^{-1} \right] \nonumber \\ 
&=&  \frac{1}{\sin^2 \phi} 
\left[ \ident -  \frac{\delta}{\sin^2 \phi} \frac{\sin^2 \phi}{\delta + \sin^2 \phi} \mt{H}_\ell \right]
=  \frac{1}{\sin^2 \phi} 
\left[ \ident -  \frac{\delta}{\delta + \sin^2 \phi} \mt{H}_\ell \right] \nonumber\\ 
&=&  
\frac{1}{\sin^2 \phi} \frac{\delta}{\delta + \sin^2 \phi} \left[ \ident -   \mt{H}_\ell \right] 
+\frac{1}{\sin^2 \phi}\frac{\sin^2 \phi}{\delta + \sin^2 \phi} \ident \nonumber\\ 
&=&  
\frac{\delta}{\sin^2 \phi (\delta + \sin^2 \phi)} \mt{P}_\ell 
+\frac{1}{\delta + \sin^2 \phi} \ident 
\end{eqnarray} 
and
$|B(\phi)| = |\sin^2 \phi ~\ident +  \delta \mt{H}_\ell| 
= (\sin^2 \phi)^n \left| ~\ident +  \frac{\delta}{\sin^2 \phi} \mt{H}_\ell \right|
= (\sin^2 \phi)^n \left| ~\mt{I}_{d_\ell} +  \frac{\delta}{\sin^2 \phi} \xtxl \invxtxl \right|$
resulting in 
$$
|B(\phi)| 
= (\sin^2 \phi)^n \left( 1+ \frac{\delta}{\sin^2 \phi}\right)^{d_\ell} 
= (\sin^2 \phi)^{n-d_\ell} ( \delta + \sin^2 \phi )^{d_\ell}. 
$$
Also
$\by^TB^{-1}(\phi) \by= 
\frac{\delta}{\sin^2 \phi (\delta + \sin^2 \phi)} \by^T\left[ \ident -   \mt{H}_\ell \right] \by
+\frac{1}{\delta + \sin^2 \phi} \by^T\by
= \frac{1}{\delta + \sin^2 \phi} \left( \frac{\delta}{\sin^2 \phi} RSS_\ell + \by^T\by \right) $.
From (\ref{aphi}), $A(\phi)$ is now given by 
\begin{eqnarray} 
A(\phi) &=& \bxo^T B^{-1}(\phi) \bxo 
 = \frac{1}{\sin^2 \phi} \bxo^T \left[ \ident -  \frac{\delta}{\delta + \sin^2 \phi} \mt{H}_\ell \right] \bxo \nonumber\\
&=&\frac{1}{\sin^2 \phi} \left[ \bxo^T\bxo -  \frac{\delta}{\delta + \sin^2 \phi} \bxo^T\mt{H}_\ell\bxo \right] 
 = \frac{1}{\sin^2 \phi} \left[ \bxo^T\bxo -  \frac{\delta}{\delta + \sin^2 \phi} \bxo^T \bxo \right]  \nonumber \\
&=& \frac{1}{\delta + \sin^2 \phi}  \bxo^T \bxo  \nonumber
\end{eqnarray} 
since $\mt{H}_\ell$ is idempotent and 
$\bxo^T \mt{H}_\ell = \bxo $ for any model $M_0$ nested in $M_\ell$.
This comes from the blockwize formula where for any 
$X_\ell=[X_0, X_{\ell\setminus0}]$ we have 
\begin{eqnarray*} 
H_\ell &=& H_0 + H_{(I_n-H_0)X_{\ell\setminus0}} \Leftrightarrow \\
X_0^T H_\ell &=& X_0^T H_0 + X_0^T H_{P_0 X_{\ell\setminus0}} \\ 
 &=& X_0^T  + X_0^T \mt{P}_0 X_{\ell\setminus0} \Big\{ [\mt{P}_0X_{\ell\setminus0}]^T \mt{P}_0X_{\ell\setminus0} \Big\}^{-1} [\mt{P}_0X_{\ell\setminus0} ]^T \\
 &=&  X_0^T  +  (X_0^T-X_0^TH_0)X_{\ell\setminus0} \Big\{ [\mt{P}_0 X_{\ell\setminus0}]^T (\mt{P}_0 X_{\ell\setminus0} \Big\}^{-1} [\mt{P}_0X_{\ell\setminus0} ]^T= X_0^T. 
\end{eqnarray*}
Therefore
$|A(\phi) |= (\delta + \sin^2 \phi)^{-d_0} | \bx_0^T \bx_0|$
and
 $\bxo A^{-1}(\phi) \bxo = (\delta + \sin^2 \phi) H_0$. 
From (\ref{ephi}) we obtain that 
\begin{eqnarray}
D(\phi) &=& \by^T B^{-1}(\phi) \by -  \by^T B^{-1}(\phi) \bxo A^{-1}(\phi) \bxo^T B^{-1}(\phi)\by \nonumber \\ 
&=& \frac{1}{\delta + \sin^2 \phi} \left( \frac{\delta}{\sin^2 \phi} RSS_\ell + \by^T\by \right)  
-  \by^T B^{-1}(\phi) [(\delta + \sin^2 \phi) \mt{H}_0] B^{-1}(\phi)\by \nonumber \\ 
&=& \frac{1}{\delta + \sin^2 \phi} \left( \frac{\delta}{\sin^2 \phi} RSS_\ell + \by^T\by \right)   -  (\delta + \sin^2 \phi)  \by^T \nonumber\\ 
&&
\left[ \frac{1}{\sin^2 \phi } \left(  \ident -   \frac{\delta}{ \delta + \sin^2 \phi }\mt{H}_\ell \right) \right]\mt{H}_0 
\left[ \frac{1}{\sin^2 \phi } \left(  \ident -   \frac{\delta}{ \delta + \sin^2 \phi }\mt{H}_\ell \right) \right] 
\by \nonumber \\ 
&=& 
\frac{1}{\delta + \sin^2 \phi} \left( \frac{\delta}{\sin^2 \phi} RSS_\ell + \by^T\by \right) \nonumber \\ && 
-  \frac{ \delta + \sin^2 \phi}{ \sin^4 \phi }  \by^T
\left(  \ident -   \frac{\delta}{ \delta + \sin^2 \phi }\mt{H}_\ell \right) \mt{H}_0 
\left(  \ident -   \frac{\delta}{ \delta + \sin^2 \phi }\mt{H}_\ell \right)  \by
\nonumber \\ 
&=& 
\frac{1}{\delta + \sin^2 \phi} \left( \frac{\delta}{\sin^2 \phi} RSS_\ell + \by^T\by \right) \nonumber \\ && 
-  \frac{ \delta + \sin^2 \phi}{ \sin^4 \phi }  \by^T
\left( \mt{H}_0  -   \frac{\delta}{ \delta + \sin^2 \phi }\mt{H}_\ell \mt{H}_0
 -   \frac{\delta}{ \delta + \sin^2 \phi }\mt{H}_0\mt{H}_\ell 
 + \left[\frac{\delta}{ \delta + \sin^2 \phi } \right]^2 \mt{H}_\ell \mt{H}_0 \mt{H}_\ell
\right) \by  \nonumber \\ 
&   \stackrel{(\mt{H}_0 \mt{H}_\ell = \mt{H}_0)}{=} & 
\frac{1}{\delta + \sin^2 \phi} \left( \frac{\delta}{\sin^2 \phi} RSS_\ell + \by^T\by \right)  
-  \frac{ \delta + \sin^2 \phi}{ \sin^4 \phi }  \left[\frac{\sin^2 \phi}{ \delta + \sin^2 \phi } \right]^2
\by^T\mt{H}_0\by \nonumber \\ 
& =  & 
\frac{1}{\delta + \sin^2 \phi}  \left( \frac{\delta}{\sin^2 \phi} RSS_\ell + \by^T\by   
-  \by^T\mt{H}_0\by \right)\nonumber \\ 
& =  & 
\frac{1}{\delta + \sin^2 \phi}  \left( \frac{\delta}{\sin^2 \phi} RSS_\ell + RSS_0 \right). \nonumber  
\end{eqnarray}

By substituting the above equations in (\ref{bf1}) we obtain 
\begin{eqnarray}
BF_{\ell \, 0}^{J-PEP} \hspace{-0.7em}
&=& \hspace{-0.7em} 2 ~
\frac{ \Gamma\left( n-d_\ell \right) }{ \Gamma\left( \frac{n-d_\ell}{2} \right)^2 } 
|\bxo^T \bxo|^{ \frac{1}{2} } RSS_0^{\frac{n-d_0}{2}}
\int \limits _{0}^{\pi/2} \ \frac{ (\sin\phi \cos \phi)^{n-d_\ell-1}  }
                                 { |B(\phi)|^{\frac{1}{2}} | A(\phi)|^{\frac{1}{2}} [D(\phi)]^{\frac{n-d_0}{2}}} d\phi  \nonumber \\ 
&=& \hspace{-0.7em}2 ~
\frac{ \Gamma\left( n-d_\ell \right) }{ \Gamma\left( \frac{n-d_\ell}{2} \right)^2 } 
|\bxo^T \bxo|^{ \frac{1}{2} } RSS_0^{\frac{n-d_0}{2}} \nonumber \\ && 
\int \limits _{0}^{\frac{\pi}{2}} \ \frac{ (\sin\phi \cos \phi)^{n-d_\ell-1}  
                                   (n + \sin^2 \phi)^{ \frac{n-d_0}{2}}   
                                    \left( \frac{n}{\sin^2 \phi} RSS_\ell + RSS_0 \right)^{-\frac{n-d_0}{2}} 
                                    }
                                 { (\sin^2 \phi)^{\frac{n-d_\ell}{2}} ( n + \sin^2 \phi )^{\frac{d_\ell}{2}} 
                                 (n + \sin^2 \phi)^{-\frac{d_0}{2}} | \bx_0^T \bx_0|^{\frac{1}{2}} } d\phi \nonumber \\ 
&=& \hspace{-0.7em}2 ~
\frac{ \Gamma\left( n-d_\ell \right) }{ \Gamma\left( \frac{n-d_\ell}{2} \right)^2 } 
\int \limits _{0}^{\frac{\pi}{2}} \ \frac{ (\sin\phi \cos \phi)^{n-d_\ell-1}  
                                   (n + \sin^2 \phi)^{ \frac{n-d_0}{2}}  (\sin^2 \phi)^{\frac{n-d_0}{2}}
                                    \left( n\frac{ RSS_\ell}{ RSS_0} +\sin^2 \phi \right)^{-\frac{n-d_0}{2}} 
                                    }
                                 { (\sin^2 \phi)^{\frac{n-d_\ell}{2}} ( n + \sin^2 \phi )^{\frac{d_\ell}{2}} 
                                 (n + \sin^2 \phi)^{-\frac{d_0}{2}}   } d\phi \nonumber \\ 
&=& \hspace{-0.7em}2 ~
\frac{ \Gamma\left( n-d_\ell \right) }{ \Gamma\left( \frac{n-d_\ell}{2} \right)^2 } 
\int \limits _{0}^{\frac{\pi}{2}} \  \frac{ (\sin\phi)^{n-d_0-1} (\cos \phi)^{n-d_\ell-1}  (n + \sin^2 \phi)^{ \frac{n-d_\ell}{2}}  }
                                  { \left( n\frac{ RSS_\ell}{ RSS_0} +\sin^2 \phi \right)^{\frac{n-d_0}{2}} }
                                  d\phi. \label{bf_final} 
\end{eqnarray}

For large $n$, we can write 
\begin{eqnarray*} 
(n + \sin^2 \phi)^{ \frac{n-d_\ell}{2}} 
&=& (n + \sin^2 \phi)^{ \frac{n}{2}} (n + \sin^2 \phi)^{ \frac{-d_\ell}{2}} 
= n^{ \frac{n}{2}}  \left(1 + \frac{\sin^2 \phi/2}{n/2}\right)^{ \frac{n}{2}} (n + \sin^2 \phi)^{ \frac{-d_\ell}{2}}  \\ 
&\approx& n^{ \frac{n}{2}}  (n + \sin^2 \phi)^{ \frac{-d_\ell}{2}} \exp \left( \frac{\sin^2 \phi}{2} \right)\\ 
&\approx& n^{ \frac{n-d_\ell}{2}}  \exp \left( \frac{\sin^2 \phi}{2} \right)~.
\end{eqnarray*} 
Similarly, 
\begin{eqnarray*}
\left( n\frac{ RSS_\ell}{ RSS_0} +\sin^2 \phi \right)^{\frac{n-d_0}{2}} 
&=& 
\left[ n\frac{ RSS_\ell}{ RSS_0} \right]^{\frac{n-d_0}{2}} 
\left( 1 + \frac{ \frac{1}{2} \sin^2\phi\frac{ RSS_0}{ RSS_\ell}}{\frac{n}{2}}  \right)^{\frac{n}{2}} 
\left( 1 + \frac{\sin^2\phi\frac{ RSS_0}{ RSS_\ell}}{n}  \right)^{\frac{-d_0}{2}} \\
&\approx& 
\left[ n\frac{ RSS_\ell}{ RSS_0} \right]^{\frac{n-d_0}{2}} 
\exp \left( \frac{1}{2} \sin^2\phi\frac{ RSS_0}{ RSS_\ell} \right). 
\end{eqnarray*}
Moreover, for large $z$ we have 
$$
\log \Gamma(z) \approx \left(z-\frac{1}{2}\right) \log z - z +\frac{1}{2} \log(2\pi).
$$
Hence 
\begin{eqnarray*}
\log \Gamma(n-d_\ell) &\approx& \left(n-d_\ell-\frac{1}{2}\right) \log (n-d_\ell) - (n-d_\ell) +\frac{1}{2} \log(2\pi) \\
\log \Gamma\left( \frac{n-d_\ell}{2} \right) 
&\approx& \left(\frac{n-d_\ell-1}{2}\right) \log \left(\frac{n-d_\ell}{2}\right) - \left(\frac{n-d_\ell}{2}\right) +\frac{1}{2} \log(2\pi) \\ 
\log \Gamma(n-d_\ell) - 2 \log \Gamma\left( \frac{n-d_\ell}{2} \right) 
&\approx&
\left(n-d_\ell-\frac{1}{2}\right) \log (n-d_\ell) - (n-d_\ell) +\frac{1}{2} \log(2\pi) \\ &&
-2\left(\frac{n-d_\ell-1}{2}\right) \log \left(\frac{n-d_\ell}{2}\right) + 2\left(\frac{n-d_\ell}{2}\right) -2\frac{1}{2} \log(2\pi) \\
&\approx&
 \frac{1}{2}  \log (n-d_\ell)  -\frac{1}{2} \log(2\pi) + (n-d_\ell-1) \log 2~ \\ 
&\approx&
 \frac{1}{2}  \log (n)  + n \log 2~. 
\end{eqnarray*}

From the above we obtain that 
\begin{eqnarray}
 \log BF_{\ell \, 0}^{J-PEP}
&\approx&   \frac{1}{2}  \log (n-d_\ell)  -\frac{1}{2} \log(2\pi) + (n-d_\ell) \log 2~ \nonumber \\ && 
+ \log 
\int \limits _{0}^{\frac{\pi}{2}} \  \frac{ (\sin\phi)^{n-d_0-1} (\cos \phi)^{n-d_\ell-1}  
                                     n^{ \frac{n-d_\ell}{2}}  \exp \left( \frac{\sin^2 \phi}{2} \right)  }
                                  { \left[ n\frac{ RSS_\ell}{ RSS_0} \right]^{\frac{n-d_0}{2}} 
                                           \exp \left( \frac{1}{2} \sin^2\phi\frac{ RSS_0}{ RSS_\ell} \right) } d\phi  \nonumber \\ 
&\approx&   \frac{1}{2}  \log (n-d_\ell)  -\frac{1}{2} \log(2\pi) + (n-d_\ell) \log 2~
+\frac{n-d_\ell}{2} \log n 
-\frac{n-d_0}{2} \log n \log 2~ \nonumber \\ && 
-\frac{n-d_0}{2} \log \frac{ RSS_\ell}{ RSS_0} 
+ \log 
\int \limits _{0}^{\frac{\pi}{2}} \  \frac{ (\sin\phi)^{n-d_0-1} (\cos \phi)^{n-d_\ell-1}   \exp \left( \frac{\sin^2 \phi}{2} \right)  }
                                  { \exp \left( \frac{1}{2} \sin^2\phi\frac{ RSS_0}{ RSS_\ell} \right) } d\phi  \nonumber \\ 
&\approx&   \frac{1}{2}  \log (n-d_\ell)  -\frac{1}{2} \log(2\pi) + (n-d_\ell) \log 2~ 
-\frac{d_\ell-d_0}{2} \log n ~ \nonumber \\ && 
-\frac{n-d_0}{2} \log \frac{ RSS_\ell}{ RSS_0} 
+ \log 
\int \limits _{0}^{\frac{\pi}{2}} \  \frac{ (\sin\phi)^{n-d_0-1} (\cos \phi)^{n-d_\ell-1}   \exp \left( \frac{\sin^2 \phi}{2} \right)  }
                                  { \exp \left( \frac{1}{2} \sin^2\phi\frac{ RSS_0}{ RSS_\ell} \right) } d\phi  \nonumber \\ 
&\approx&   \frac{1}{2}  \log n + n \log 2~ 
-\frac{d_\ell-d_0}{2} \log n -  \frac{n}{2} \log \frac{ RSS_\ell}{ RSS_0}  \nonumber \\ 
\end{eqnarray}
since the integral 
$$
\int \limits _{0}^{\frac{\pi}{2}} \  \frac{ (\sin\phi)^{n-d_0-1} (\cos \phi)^{n-d_\ell-1}   \exp \left( \frac{\sin^2 \phi}{2} \right)  }
                                  { \exp \left( \frac{1}{2} \sin^2\phi\frac{ RSS_0}{ RSS_\ell} \right) } d\phi  
                                  \le 
\int \limits _{0}^{\frac{\pi}{2}} \     \exp \left( \frac{\sin^2 \phi}{2} \left[1-\frac{ RSS_0}{ RSS_\ell} \right] \right)  d\phi 
$$
when $n \ge d_0+1$ and $n \ge d_\ell+1$. 
The latter integral has a finite value for all $n$ according to \citet[p.1216]{casella_etal_2009}. 
Hence the integral involved  in the $BF_{\ell \, 0}^{J-PEP}$ has also a finite value for all $n$. 

If we compare any two models $M_\ell$ and $M_k$ (both of them different than the reference model) 
we have that 
\begin{eqnarray}
\label{cons}
 -2 \log BF_{\ell \, k}^{J-PEP}
&\approx&   n \log \frac{ RSS_\ell}{ RSS_k} + (d_\ell-d_k) \log n
=  BIC_\ell - BIC_k ~.
\end{eqnarray}

Therefore the J-PEP approach has the same asymptotic behavior as the
BIC-based variable-selection procedure. 
The following Lemma is a direct result of (\ref{cons}) and of Theorem 4 of \cite{casella_etal_2009}.

\begin{lemma} 
\label{lemma 1}

Let $M_\ell \in {\cal M}$  be a normal regression model of type (\ref{new2-1}) such that 
$$
\lim \limits _{n\rightarrow \infty} \frac{ \mt{X}_T \big( \mt{I}_n -\mt{X}_\ell(\mt{X}_\ell^T \, \mt{X}_\ell)^{-1} \mt{X}_\ell^T \big) \mt{X}_T }{n} \mbox{ is a positive semidefinite matrix}, 
$$
with $X_T$ being the  design matrix  of the true data generating regression model  $M_T \neq M_\ell$. 
Then, the variable selection procedure based on J-PEP Bayes factor is consistent since
$BF^{J-PEP}_{\ell T} \rightarrow 0$ as $n\rightarrow \infty$.

\end{lemma}

\section{Simulation Study}

In this section, we perform a simulation comparison that studies the behavior of the proposed method 
as the sample size increases. We compare the performance of our method with that  
of the ``most established" Bayesian variable selection techniques: the $g$-prior \citep{zellner_76}, the hyper-$g$ prior
\citep{liang_etal_2008}, the \cite{Zellner_siow_80} prior and the BIC \citep{schwarz_78}. All competing methods were implemented using  
the \texttt{BAS} package in \texttt{R}; we set $g = n$ in the $g$-prior to
correspond to the unit information prior \citep{kass_wasserman_95} and $\alpha = 3$ in
the hyper-$g$ prior as recommended by
\cite{liang_etal_2008}. For the implementation of our approach we used the second Monte Carlo scheme presented in Section 3 of \cite{fouskakis_et.al_2014}.

We consider 100 simulated data-sets of sample sizes $n$ = 30, 50, 100, 500, 1000 and $p=10$ covariates generated from a standardized Normal distribution, while the response is generated from
\begin{eqnarray}
\label{ss}
 Y_i \sim N(0.3X_{i3} + 0.5X_{i4} + X_{i5}, 2.5^2), \quad \mbox{for} \quad i=1,\dots,n.
\end{eqnarray}

\begin{figure}[h!]
\caption{ Boxplots (per 100 simulated datasets of different sample sizes) of the posterior probability of the true model for different variable selection methods.}
\label{true_model}
\begin{center}
\psfrag{Hyper-g alpha=3}[l][l][0.65]{{\sf Hyper-g ($\alpha=3$)}}
\includegraphics[scale=0.55, angle=-90]{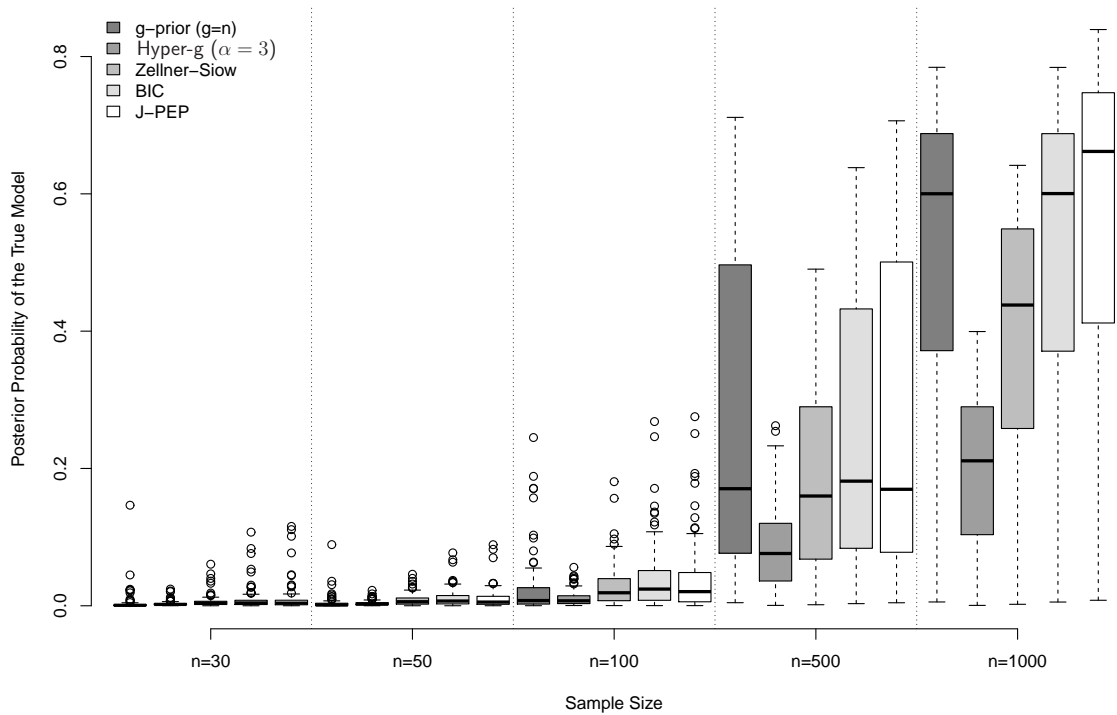}
\end{center}
\end{figure}

Figure \ref{true_model} depicts the between-samples distribution of the posterior probability of the true model for the Bayesian variable selection techniques 
under comparison.
It is clear that for small sample sizes all competitive methods fail to provide high posterior evidence in favor of the true model. 
As the sample size gets larger, all methods increase their posterior support towards the true model, 
with  the proposed J-PEP method to perform slightly better than the Zellner's g-prior and the BIC. 
This is sensible since these three methods are converging to the same Bayes factors as $n$ grows but 
with J-PEP constantly supporting more parsimonious models. 
On the other hand, the hyper-g prior gives the lowest support towards the true model 
due to its hierarchical structure which increases the posterior uncertainty on the model space. 
Practically, the hyper-g prior needs larger sample size, than the rest of the methods, in order 
to fully a-posteriori support the true generating mechanism.

\begin{figure}[p!]
\caption{ Boxplots (per 100 simulated datasets of different sample sizes) of posterior inclusion probabilities for each covariate under the different variable selection methods.}
\label{inc_probs}
\begin{center}
\psfrag{Hyper-g \(a=3\)}[l][l][0.65]{{\sf Hyper-g  ($\alpha=3$)~~}}
\includegraphics[scale=0.85]{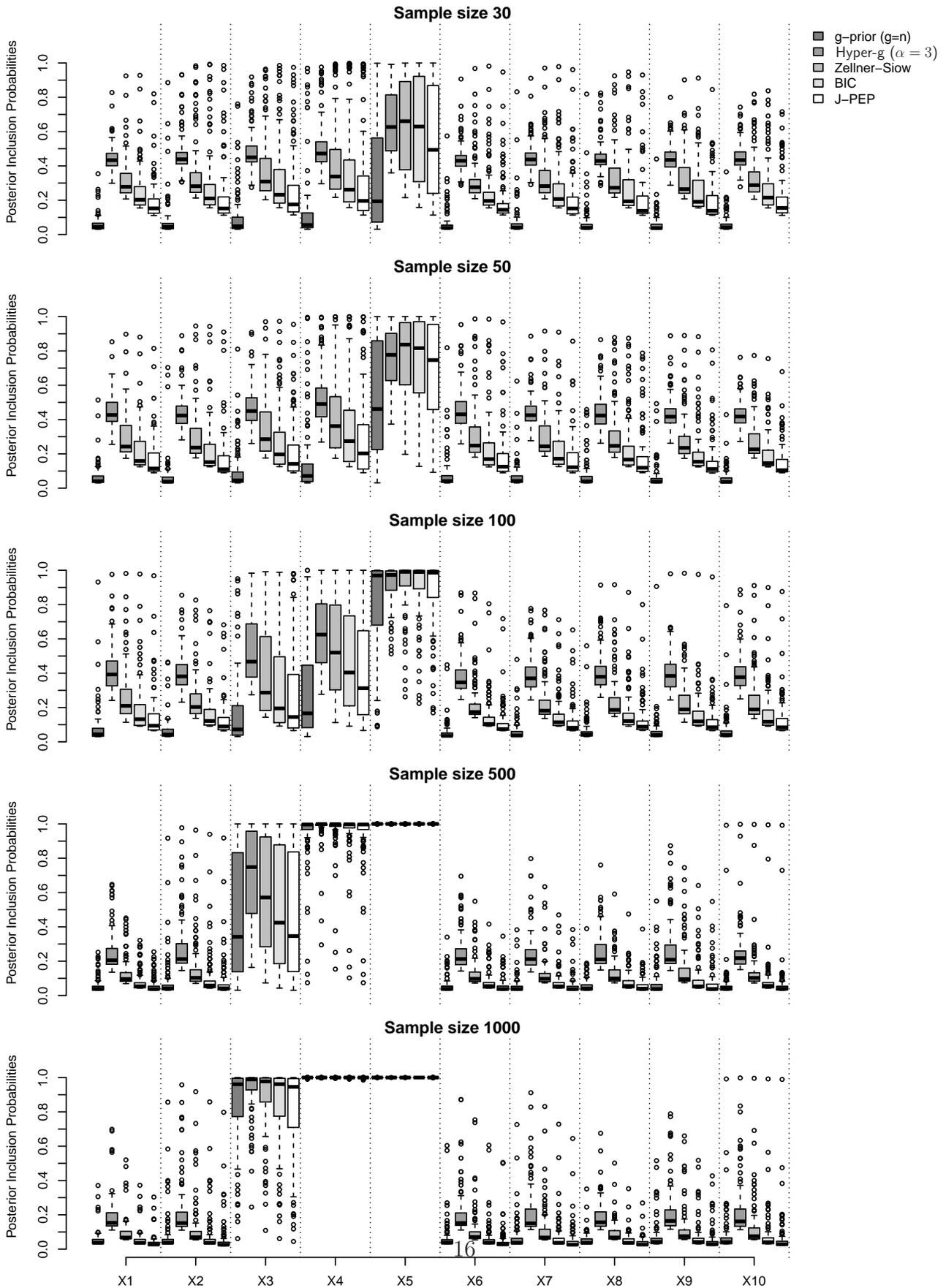}
\end{center}
\end{figure}

Looking now at the posterior inclusion probabilities of each covariate in Figure \ref{inc_probs}, 
we observe that all methods successfully identify $X_5$ (with true effect equal to one) as an important 
component of the model, even for small sample sizes, with the exception of the Zellner's g-prior. 
Furthermore, the between-samples variability of the posterior inclusion probabilities reduces as the sample size increases. 
Returning back to the Zellner's g-prior, it  fails to a-posteriori support $X_5$ for $n=30$ and $n=50$.  
Generally, the g-prior demonstrates much larger between-sample variability than the rest of the methods 
and it seems to be unable to identify the true effects for small sample sizes in this simulation study. 

Similar is the picture for the posterior inclusion probabilities of the other two covariates with non-zero effects, $X_3$ and $X_4$, 
but with slower rates of convergence towards to one. 
For the latter covariate (with true effect equal to $0.5$) we observe large between-samples 
uncertainty concerning the importance of this effect for $n \le 100$ under all methods. 
For $n\ge 500$, all methods successfully identify the importance of this covariate with small between-samples variability. 
In general, the hyper-g method supports this covariate with the highest inclusion probabilities 
while the J-PEP with the lowest inclusion probabilities. 
This is  due to the characteristics of the two methods, with the first supporting more complicated models 
while the latter more parsimonious ones. 
We reach to similar conclusions for  covariate $X_3$ (with true effect equal to $0.3$) 
but with the addition that the Zellner's g-prior does not spot the effect of this covariate as important, even for samples of size $n=500$. 
Moreover, we need to increase the sample size to $n=1000$, for all methods, in order to obtain high posterior inclusion probabilities 
with relatively low between-samples variability. 

Reasonably, the between-samples distribution of the posterior inclusion probabilities is similar for all covariates with zero true effects. 
It is noticeable that all methods, except the hyper-g prior, 
identify, really fast, that these covariates should have low posterior inclusion probabilities with the 
between-samples variability considerably to decrease as $n$ gets larger. 
On the other hand, the posterior inclusion probabilities under the hyper-g prior setup are systematically higher (close to $0.5$)
than the corresponding ones under the other competing methods. 
This increases the posterior uncertainty on the model space and results 
to lower probabilities of identifying the true model as the maximum a-posteriori model. 
It is also noticeable that these posterior inclusion probabilities, under the hyper-g prior setup, 
both in terms of median values and in terms of between-samples variability, 
seem to converge very slowly towards zero as $n$ gets larger. 

To sum up, in this simulation study the J-PEP prior methodology identifies the true model structure with (slightly) higher posterior probability than the rest of the methods. It provides posterior inclusion probabilities close to zero for non-important effects (even for small sample sizes) and high inclusion probabilities for the important effects (although these are smaller than the ones obtained under the competing methods for small sample sizes).

\section{Discussion}

Under the \textit{power-expected-posterior prior} (PEP) approach, ideas from the power-prior and unit-information-prior methodologies are combined. As a result the PEP priors are minimally-informative and the effect of training samples is 
reduced. When using the independence Jeffreys as a baseline prior for normal linear models, we prove that the J-PEP approach has the same asymptotic behavior as the BIC-based variable-selection procedure. Therefore, under very mild conditions on the design matrix, it is a consistent variable selection technique. 

\section*{Acknowledgments} 
  We wish to thank the Editor and the referee for comments that greatly strengthened the paper. This research has been co-financed in part by the European Union (European Social Fund-ESF)
and by Greek national funds through the Operational Program ``Education and
Lifelong Learning" of the National Strategic Reference Framework
(NSRF)-Research Funding Program: Aristeia II/PEP-BVS.

\bibliographystyle{agsm}

\bibliography{biblio3}

\end{document}